\documentclass[%
 aip,
rsi,%
 amsmath,amssymb,
 reprint,%
floatfix
]{revtex4-1}

\usepackage{graphicx}
\usepackage{dcolumn}
\usepackage{bm}

\usepackage{mciteplus}
\usepackage[version=4]{mhchem}
\usepackage[capitalise]{cleveref}
\usepackage{booktabs}
\usepackage{multirow}
\usepackage{subcaption}
\usepackage[section]{placeins}
\usepackage{bm}

\begin{document}

\preprint{AIP/123-QED}

\title[Neural evolution structure generation]{Neural evolution structure generation: High Entropy Alloys}

\author{Conrard Giresse Tetsassi Feugmo}
 \email{ConrardGiresse.TetsassiFeugmo@nrc-cnrc.gc.ca}
 \affiliation{National Research Council Canada}
 \author{Kevin Ryczko}
 \affiliation{Department of Physics , University of Ottawa}
  \affiliation{Vector Institute for Artificial Intelligence, Toronto, Ontario, Canada }
  \author{Abu Anand}
 \affiliation{Department of Materials Science and Engineering , University of Toronto}
  \author{Chandra Veer Singh}

 \affiliation{Department of Materials Science and Engineering , University of Toronto}
 \affiliation{Department of Mechanical and Industrial Engineering, University of Toronto}
\author{Isaac Tamblyn}%
\email{isaac.tamblyn@nrc.ca}
 \affiliation{National Research Council Canada}
 \affiliation{Department of Physics , University of Ottawa}
 \affiliation{Vector Institute for Artificial Intelligence, Toronto, Ontario, Canada }


\date{\today}

\begin{abstract}
We propose a method of neural evolution structures (NESs)  combining artificial neural networks (ANNs) and evolutionary algorithms (EAs) to generate  High Entropy Alloys (HEAs) structures. Our inverse design approach is based on pair distribution functions and atomic properties and allows one to train a model on smaller unit cells and then generate a larger cell. With a speed-up factor of  approximately  1000 with respect to the SQSs, the NESs dramatically reduces computational costs and time, making possible the generation of very large structures (over 40,000 atoms) in few hours. Additionally, unlike the SQSs, the same model can be used to generate multiple structures with same fractional composition.
\end{abstract}

\maketitle

\section{Introduction}


Multicomponent alloy systems such as High Entropy Alloys (HEAs), and Bulk Metallic Glasses (BMGs) have been in the physical metallurgy research spotlight over the past decade \cite{kruzic2016bulk, tsai2014high}. 
HEAs are particularly interesting because of their superior structural and functional properties \cite{tsai2014high,Sahlberg2016, Amiri2021, Nyby2021, Pickering2021}. In contrast to the conventional notion of alloying with a principal element (solvent) and alloying elements (solute), HEAs have four or more principal elements in near-equiatomic compositions \cite{YE2016, Zhou2018, IKEDA2019464, george2019high}. 

Computational modeling is necessary for targeted and rapid HEAs discovery and application\cite{Wan2013, LI2017262, Rogal2017, Yoav2018364, Hu2021, Strother2021}. Constructing an appropriate atomic structure is the first step towards reliable predictions of materials properties. This includes predicting thermodynamic, kinetic, electronic, vibrational, and magnetic properties \cite{Duancheng201590, Kormann2015, Troppenz2017, chattaraj2018structural, chen2019diffusion, zhao2018novel, rostami2020bulk}, with first-principles methods based simulation methodologies like Density Functional Theory (DFT). Indeed, DFT modeling of complex, random alloys requires defining a fixed-size cell \cite{Gao2013, Wang2019, Yin2019, Hu2020}, which can introduce non-random periodicity. The inherent local disorder of HEAs makes this a non-trivial task\cite{gao2016applications,gao2017computational,ren2019swamps}.   Special quasi-random structures (SQSs) designed to approximate the radial distribution function of a random \cite{zunger1990special,wei1990electronic} system is a quintessential concept to generate realistic random structures when modeling disordered alloys with atomic resolution.  Modern SQS generation approaches utilize techniques such as cluster expansions (CEs) in combination with Monte Carlo (MC) algorithms. Several codes are available in the literature including ICET\cite{ICET}, ATAT MCSQS\cite{VanDeWalle2013}, \textit{Supercell}\cite{okhotnikov2016supercell} which can generate SQS structures for multi-component systems. Although very powerful, these approaches have significant computational overhead. A detailed analysis of computing time with ICET with the number of atoms is presented in the subsequent sections. Along with the computational complexity, present SQS generating techniques require the optimization of multiple parameters, including, but not limited to: cluster space cutoffs, number of optimization steps, and simulated annealing temperatures for each system \cite{ICET,VanDeWalle2013}.  These create a serious bottleneck in exploring multi-component alloy systems using first-principles simulations and molecular dynamics.


An alternative is to use machine learning models to achieve desired property \cite{WEN2019109, Chen2020, Kaufmann2020, Rickman2020}. Recent works have used surrogate models \cite{Ryan2018, Liang2020}, evolutionary algorithms \cite{Oganov:2010,Oganov:2018, Podryabinkin2019}, and generative adversarial networks \cite{Sungwon:2020, Chen2020} to predict crystal structures. The inverse design framework that combine artificial networks and evolutionary algorithms have also had success \cite{Zunger2018, Ryczko:2020} in generating structures that optimize some objective function. The generated structures can be used to collect descriptors such as structural stability, lattice vibrational property, electronic structure, elasticity, and stacking fault energy. 


In this work, we build on previous work and present a  neural evolution structures (NESs) generation algorithm that combines artificial neural networks (ANNs) and evolutionary algorithms (EAs) to enable the search of HEAs that optimizes configurational entropy. In \cref{methods}, we outline our methodology including the general workflow of the algorithm, the crystal representation, and the fitness (or objective) functions. In \cref{results}, we present our results. This includes a comparison of our algorithm with SQS with respect to performance and timing, and analysis of the optimization parameters. In \cref{conclusion}, we summarize and propose future work based on our findings.  



\section{ Methods}
\label{methods}
\subsection{General workflow}
\begin{figure*}[hbt!]
\centering
\includegraphics[width=1\linewidth]{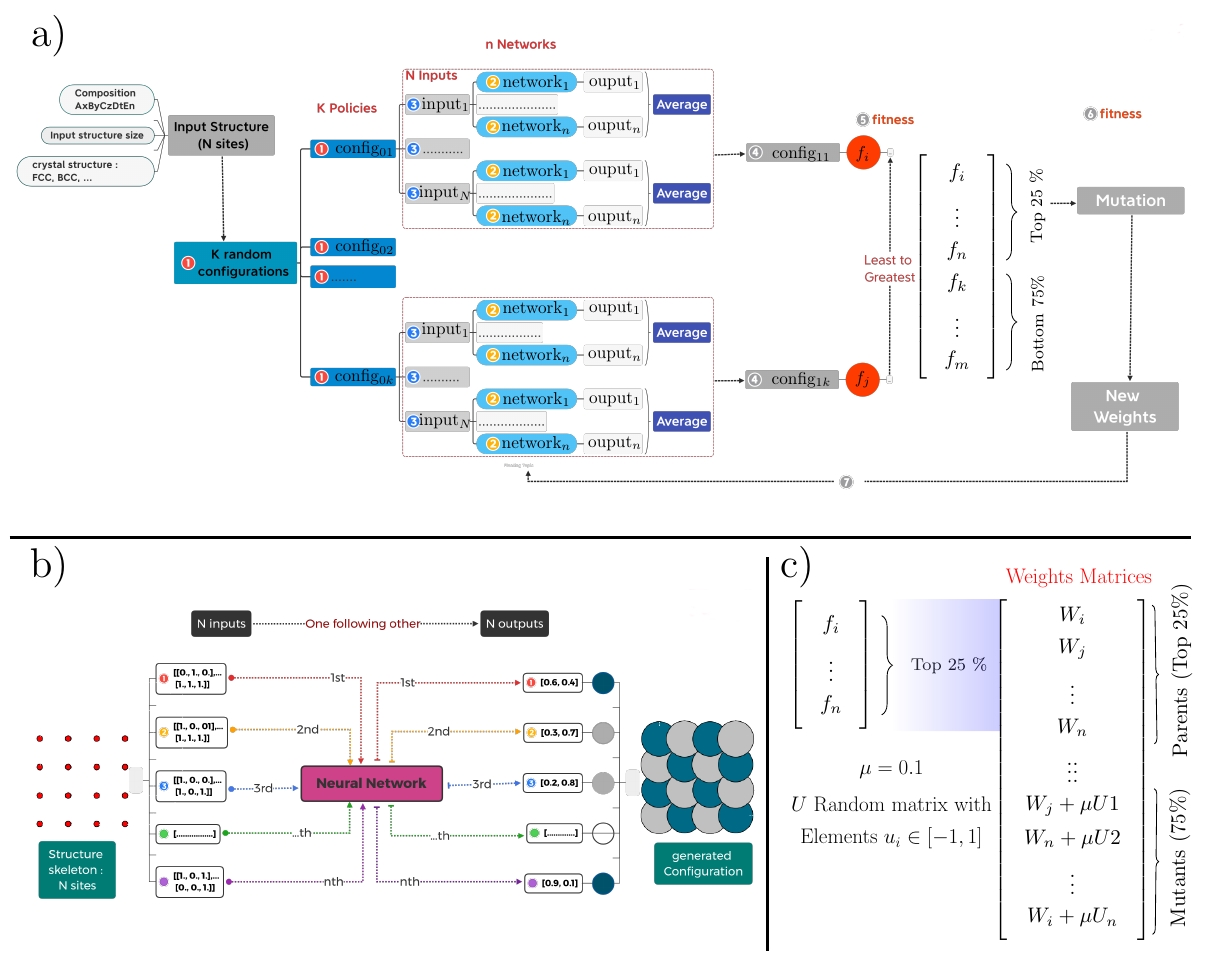}  
	\caption{Sketch of the algorithms required for the NESs generation. a) Key steps of the NES training process. b) Road-map of the NESs generation process. From left to right: an input structure (a template mesh with N sites),  a representation of the N input arrays, the ANNs, a representation of the N output vectors,  atom-type associated to each output vector,  and the generated configuration. c) Sketch of the mutation step.  W$_i$ is a weights matrix and  25\% represent  the percentage of the matrix elements mutated}
	\label{fig:workflow}
\end{figure*}

In this work, we search for HEA structures that maximize the physical disorder, or the maximum-entropy configuration. To do so, we consider configurations of HEAs that contain $M$ different atomic species over $N$ lattice sites.  The workflow is summarized in Figure \cref{fig:workflow}, which combines ANNs and EA. This methodology was used to optimize the doping of graphene-based three-terminal devices for valleytronic applications  \cite{Ryczko:2020}. This workflow can be divided into two processes: the training process (\cref{fig:workflow}-a) and the generation process (\cref{fig:workflow}-b). After defining the crystal structure (FCC, BCC, HCP, etc.), the fractional composition (\ce{A_{\alpha}B_{\beta}C_{\gamma}D_{\zeta}E_{\eta}}), and the size of the supercell, the algorithm is as follows:
\begin{enumerate}
    \item Place atoms randomly on the lattice following the fractional composition for $N$ structure copies.
    \item Initialize $M$ ANN policies for each structure copy. One could have 1 ANN policy per structure or average over many ANN policies.
    \item Generate input arrays (one input array per lattice site) based on the local environment of the lattice sites and feed the vectors into the ANN policies.
    \item Based on the outputs of the ANN policies, reassign atoms to sites in the crystal for each structure copy.
    \item Calculate the fitness function across each structure copy, sort from least to greatest, and order the associated structures.
    \item Select the top 25\% best-performing structures, and randomly select and mutate the weights of the ANN policy to generate the remaining 75\% of the population.
    \item Go to step 2.
\end{enumerate}

 The training process is repeated until convergence is reached. The input vector is based on the environment of a lattice. The first element is a vector describing the properties of the atom (i.e. the atomic number, valence electrons, etc.) for the selected lattice site. The remainder of the input vector comes from the concatenation of the atomic properties of the neighboring and next-nearest neighboring lattice sites. The input vector, therefore, changes based on the lattice structure. The final input vector is flattened such that it can be passed into the ANN policy. 
  We used the \texttt{Softmax} activation function to convert the ANN output vector into a vector of probabilities of assigning a certain chemical element to a certain lattice site. The index with the highest probability is extracted and matched to the list of elements (`A', `B', `C', `D', `E', $\ldots$) and the corresponding element is assigned to the considered site of the training structure (step 6). 
 
 Steps 3 to 6 are iterated over the remaining lattice sites until the new configuration is generated (\cref{fig:workflow}-b). For each structure, $M$ different policies are created, then for each ANN policy $N$ configurations are generated and the corresponding fitness functions are computed (step 7). Finally, the average fitness of each policy is evaluated and the averages are sorted from least to greatest. 
 
The top performing 25\% ANN policies are kept, and the rest of the population are eliminated. To reproduce the next generation, the ANN policies from the top performers are randomly selected the weights are cloned  and randomly mutated to generate the remaining 75\% (\cref{fig:workflow}-c). The new atomic configurations are generated and the process is repeated. The random mutations consist of adding a random matrix to the parents' weight (\cref{eq:mutation}).  

\begin{align}
	\label{eq:mutation}
	\begin{pmatrix}
	w_1\\ w_2\\\vdots \\w_l
	\end{pmatrix}_{new}	
	= 
	\begin{pmatrix}
	w_1\\ w_2\\\vdots \\w_l
	\end{pmatrix}_{old}	
	+ \mu \cdot 
	\begin{pmatrix}
	u_1\\ u_2\\\vdots \\u_l
	\end{pmatrix}	
\end{align}
where $\mu = 0.1$ is a small parameter (similar to a learning rate), $u_i \in [-1, 1]$ is a random number, and $l$ is the number of weights. In this work, the number of weights is equal to the number of  elements in an input array. All calculations in this report were run using 1 CPU and 8Gb RAM on an HP Z4 G4 Workstation (Intel Xeon).

\subsection{Representation of crystal structures}

We now describe the generation of the input vectors for each lattice site. The central idea of our approach is to use the  pair distribution function (PDF) to characterize a crystal structure. Indeed, for crystals, the  number of nearest neighbours, and their positions  depend on the crystal structure, and the lattice   parameters, respectively. In the FCC structure, each atom has 12  nearest neighbours (coordination number) at a distance  $d=a\sqrt{2}/2 $, 6 at $ d=a $, and 24 at  $d=a\sqrt{3}/\sqrt{2} $. In the BCC structure, each atom has 8  nearest neighbours at a distance $d=a\sqrt{3}/2 $, 6 at $ d=a $, and 12 at  $d=a\sqrt{2} $. 

The number of $a$-type atoms around an $b$-type atom is given by 
\begin{align}
\label{eq:N_ab}
N_{a b} (r_{min}, r_{max} ) 
&= 4 \pi c_{b} \rho_0 \int_{r_{min}}^{r_{max}} r^2 g_{a b}(r) dr
\end{align} 
where $r_{min}$ and $ r_{max}$  are two the radii values between which the
coordination number is to be calculated, and $c_{b}$ is the fractional composition of $b$. The partial PDF $g_{a b}(r)$ between types of particles $a$ and $b$	 reads

\begin{align}
\label{eq:g_ab}
g_{ab}(r) =   \dfrac{N}{\rho_0 N_a N_b}   \sum_{i=1}^{N_a}  \sum_{j=1}^{N_b}  \left\langle \delta (|r_i-r_j| -r)  \right\rangle
\end{align}
where $\delta$ is a Dirac $\delta$-function, and $\rho_0= N/V$ is the average density.

\begin{figure}[hbt!]
	\centering
	\includegraphics[width=1\linewidth]{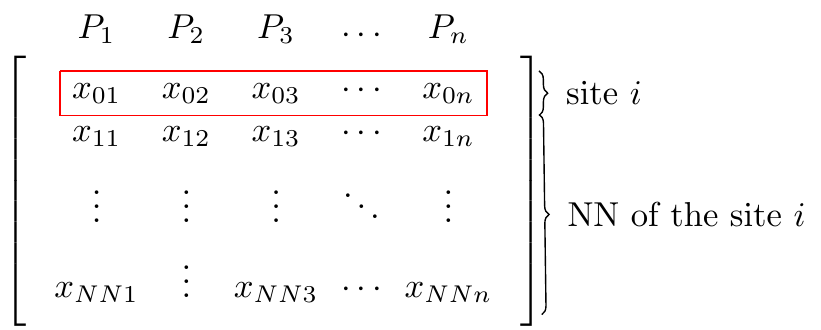}
	\caption{Sketch of the NES  input-vector for  a single crystal site.The $P_i$ column corresponds to the  $i$-property and NN is the number of nearest neighbours}
	\label{fig:inputvectors}
\end{figure}

Each crystal site is represented by an array in which the elements are the atomic properties of the chemical element occupying the site and those of its nearest neighbors (\cref{fig:inputvectors}).
The number of rows corresponds to the number of nearest neighbors (NN) plus one (NN +1) and the number of columns is equal to the number of atomic properties describing each chemical element ($P_1, P_2,  \cdots,  P_n$). The number of input-vectors is equal to the number of site in the crystal structure. The properties of the chemical element occupying the $i^{th}$-site are always stored in the first row of the $i^{th}$-input-vector. These atomic properties can be classified into quantitative and qualitative variables. The quantitative variables include the atomic number, the number of valence electrons, the electronegativity, oxidation state, and atomic radius. The qualitative variables include the row and the group (metal, transition metal, alkali, alkali, and metalloid) in the periodic table. They are represented by integer and boolean numbers, respectively. 

\subsection{Fitness functions}

\begin{figure}[hbt!]
	\centering
	\includegraphics[width=0.4\linewidth]{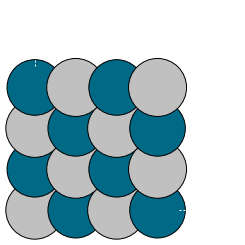}
	\includegraphics[width=0.4\linewidth]{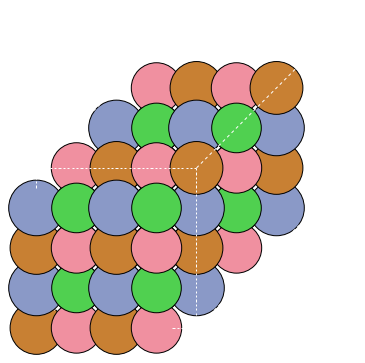}\\
	\includegraphics[width=0.2\linewidth]{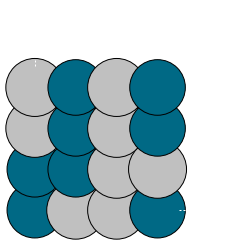}
	\includegraphics[width=0.2\linewidth]{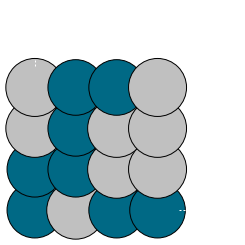}
	\includegraphics[width=0.2\linewidth]{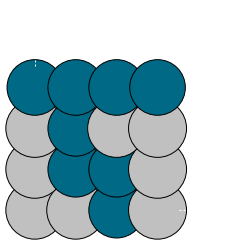}
	\includegraphics[width=0.2\linewidth]{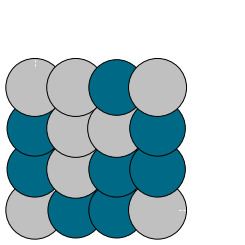}
	\caption{Example of  configurations for an equiatomic  2- and 4-components systems. Top: High entropy configurations. Bottom: Random configurations }
	\label{fig:configurations}
\end{figure}

The fitness function describes the quality of a configuration  (high entropy $\equiv$ physical disorder), and determines the best solution. Our objective is to reduce the segregation of chemical elements or to maximize the entropy of the configuration. Examples of 2- and 4- equiatomic high entropy configurations are presented in \cref{fig:configurations} (top row) along with 2- equiatomic random configuration (bottom row). Characteristic of  4-components high entropy  have been studied using a  $4\times 4 \times 4$ supercell (64 atoms), and four functions characterizing the disorder in the  crystal structures have been derived.  For a site occupied by an A-type atom: 

\begin{itemize}
	\item The first fitness function minimizes the number of A-type occupying the nearest neighbours site in the first coordination shell $N_{aa}$. Knowing that the target is 0, the fitness defined as the root-mean-square deviation from $0$
	\begin{align}
	\label{eq:fitness_AA1}
	F_{AA}^{1} = \sum_{a}\sqrt{\dfrac{\sum_{i=1}^{N_a}  \left(N_{aa}^i -0 \right)^2 }{N_a}}
	\end{align}
		\item The second fitness function minimizes the number of A-type occupying the nearest neighbours site in the second coordination shell $N_{aa}$.  If  NN$_{2}$ is the number of nearest neighbours in the second coordination shell then the fitness function reads 
		
	\begin{align}
	\label{eq:fitness_AA2}
F_{AA}^{2} = \sum_{a}\sqrt{\dfrac{\sum_{i=1}^{N_a}  \left(N_{aa}^i -NN_2 \right)^2 }{N_a}}
	\end{align}
	\item The third fitness function  equalizes the number of other types of  atoms occupying the nearest neighbour site in the first coordination shell $N_{ab}$, and reads
		\begin{align}
	\label{eq:fitness_AB}
F_{AB} =  \sum_{a \neq b} \sqrt{\dfrac{\sum_{i=1}^{N_b}  \left[N_{ab}^i -(c \cdot NN_1 ) \right]^2 }{N_b}},
	\end{align} where $c = c_{a}+\left[  c_{b}/(s-1)\right] $. $c_{a}$ and $c_{b}$ are the target  fractional compositions of $a$ and b, respectively, and $s$ is the number of atom-types. NN$_{1}$  is the number of nearest neighbours in the first coordination shell

	\item The last  fitness function  checks how the  maximum number of each types of atoms ($N_a$, $N_b$,....) deviate from the target composition. These numbers are proportional to the fractional composition 
		\begin{align}
	\label{eq:fitness_N}
	F_{N} =  \sqrt{\dfrac{\sum_{a} \left[N_a - (c_a\cdot N) \right]^2 }{ s}}
	\end{align}

\end{itemize}

The minimum of the total fitness [\cref{eq:fitness}] depends on both the fractional composition and the number of components, and it is not necessarily equal to 0. As an example, for a 2-components system, the minimum will never be equal to 0.

\begin{align}
	\label{eq:fitness}
F =  F_{AA}^{1} + F_{AA}^{2} +F_{AB} +F_N
	\end{align}

\section{Results and Discussion}
\label{results}
\subsection{NESs computation time}
\begin{figure}[hbt!]
	\centering
	\includegraphics[width=0.7\linewidth]{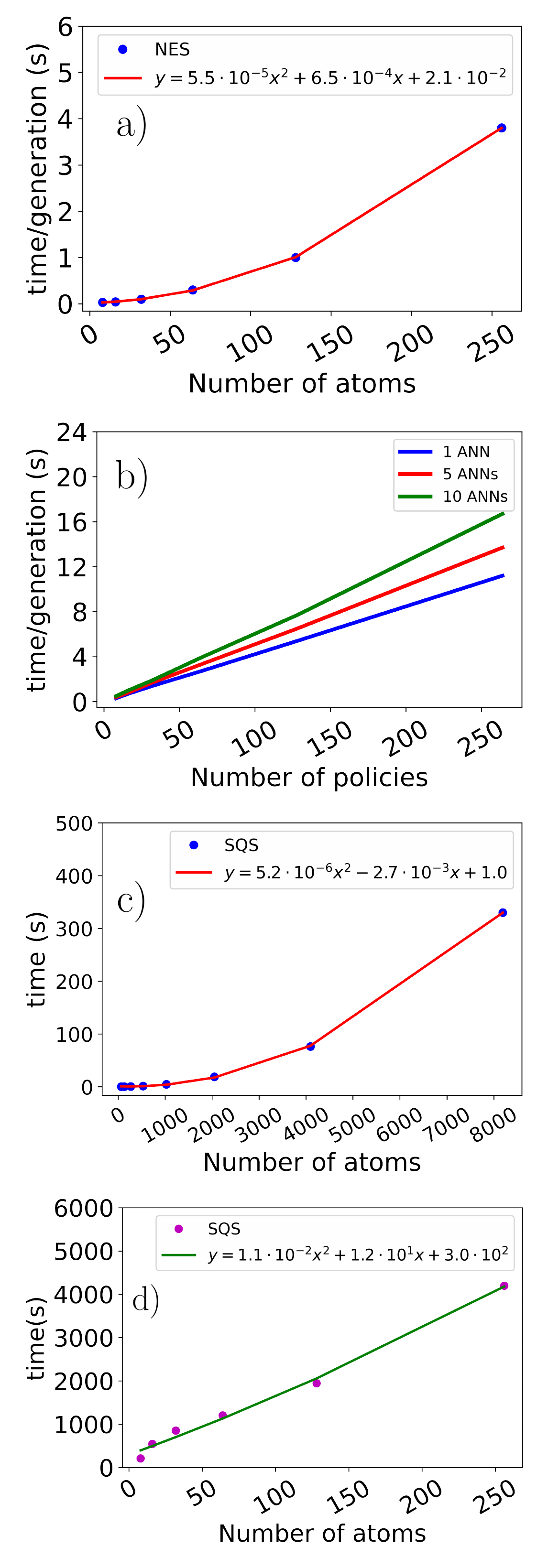}
	\caption{NES  computing time.   a) Average training time vs the size of input-structure. Calculations were carried out using 1 policy made with 1 ANN b) Average training time per generation vs the numbers of policies. Calculations were carried out using a structure of 64 atoms. c) Average NESs generation time as a function of the size of the system. d) ICET-SQSs generation time as a function  of the size of the system }
	\label{fig:times}
\end{figure}

All  calculations were carried out on equiatomic  \ce{Cu_{\alpha}Ni_{\beta}Co_{\gamma}Cr_{\zeta} }  FCC  alloy structures.
Important aspects of the algorithm are the optimization and   generation times which depends on three parameters: i) the number of policies optimized simultaneously ii) the size of  the input-structures 
and ii) the number of ANN included in each policy. The three parameters have been investigated and the results are presented in \cref{fig:times}. 

First, the average training time per generation as a  function of the size of the input-structure is shown in \cref{fig:times}-a. This figure shows that the training time increases slowly with the size of the input-structure (ratio of 1.4).

Second, \cref{fig:times}-b shows the average training time per generation as a  function of the number of policies optimized. One observes a linear increase (r = 0.99). In addition,  the slope also increases with the number of ANNs reaching  0.04, 0.05, and 0.06 for 1, 5, and 10 ANNs, respectively.

Finally, the average time per generation as a function of the size of the input-structure is shown in  \cref{fig:times}-c. It increases slowly with the size of the structures, going from few  a tenths of a second (up to 256 atoms) to few a hundred seconds around 8000 atoms ( ratio of 0.3) 

Moreover, by comparing \cref{fig:times}-a to \cref{fig:times}-d, we observe that the  SQSs scale up linearly [\cref{eq:sqs}] with  time, whereas the NESs follow a  $x^{2}$ polynomial behaviour ( but with a very low value of $a$ and $b$) [\cref{eq:nes}]. Knowing that the number of steps towards convergence increases with the increase in the number of atoms, and  taking into account that the NE is trained on small clusters, it will always require fewer steps.  Additionally, NESs can be sped-up with multiprocessing.  Taking into account all the previous remarks, we can derive a speed-up factor of approximately  1000 by comparing the  $x^{2}$ coefficients in  \cref{eq:sqs,eq:nes}. 
{\small
\begin{align}
\text{SQSs:} & &y&=1.1\cdot10^{-2} x^{2} +1.2\cdot10^{1} x+ 3.0 \cdot10^{2} \label{eq:sqs}\\
\text{NESs:} & &y&=5.5\cdot10^{-5} x^{2} + 6.5\cdot10^{-4} x+ 2.1\cdot10^{-2}\label{eq:nes}
\end{align}}

\subsection{Convergence of NESs}

\begin{figure}[hbt!]
	\centering
	\includegraphics[width=0.7\linewidth]{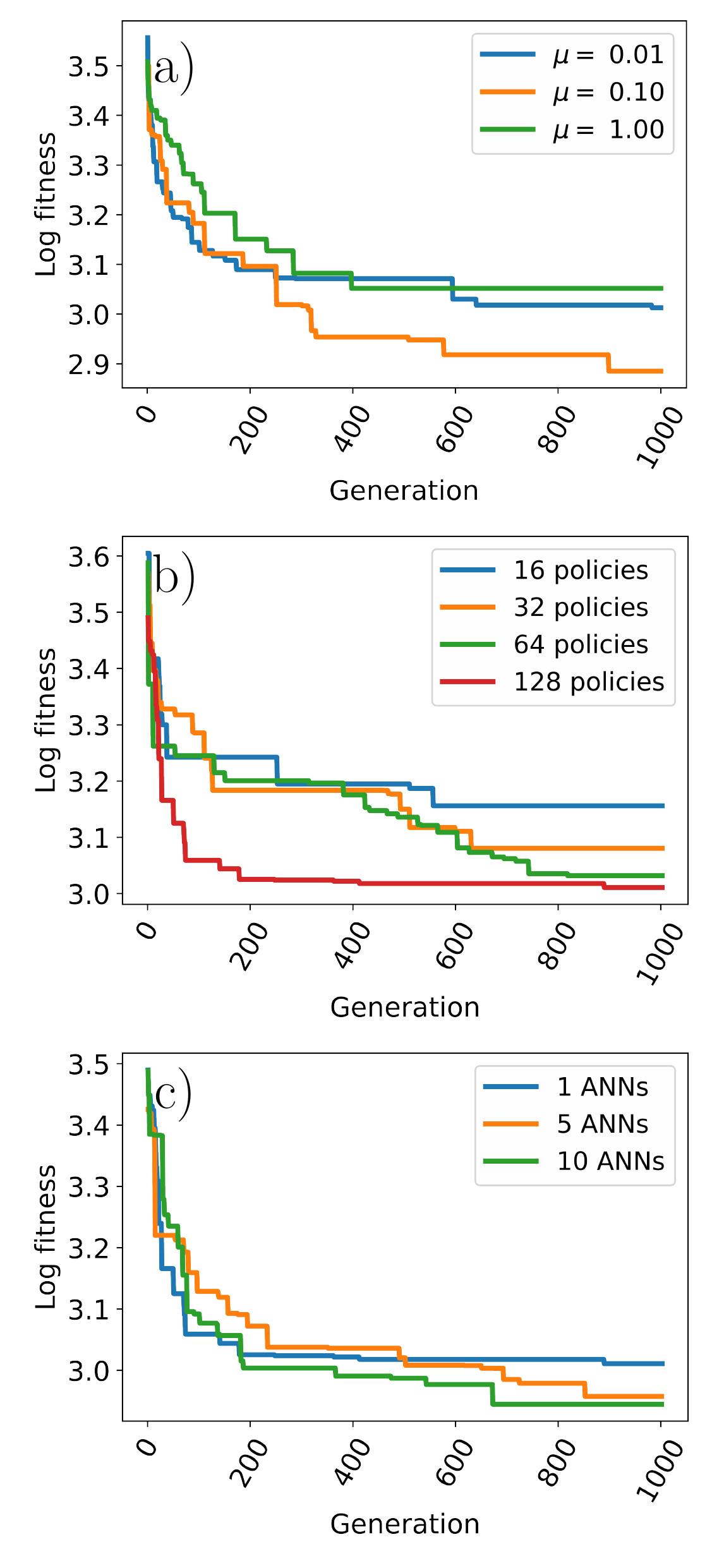}

	\caption{NES training curves. a) Optimization of the scaling factor $\mu$.
	b)Optimization of the of number policies trained simultaneously. c) Optimization of the  the number of ANNs considered in each policy. Calculations  were carried out on input-structure of 64 atoms, and for an equiatomic \ce{Cu_{\alpha}Ni_{\beta}Co_{\gamma}Cr_{\zeta} }}
	\label{fig:optimization}
\end{figure}

In addition to the computation time, we now analyze the convergence of our algorithm. We show results of optimizations with different parameters in \cref{fig:optimization}. These parameters include $\mu$  (\cref{fig:optimization}-a), the number of policies trained (\cref{fig:optimization}-b), and number of ANNs considered in each policy (\cref{fig:optimization}-c). Firstly, we investigated three values of  $\mu$ have (0.01, 0.1, and  1) and find that the best convergence is reached with 0.1. Second, the increase in the number of policies considered accelerates the convergence. Finally, in \cref{fig:optimization}-c, we see that increasing the number of ANNs per policy does not improve the convergence rate but improves the quality of the solution.  

\begin{figure}[hbt!]
	\centering
	\includegraphics[width=0.7\linewidth]{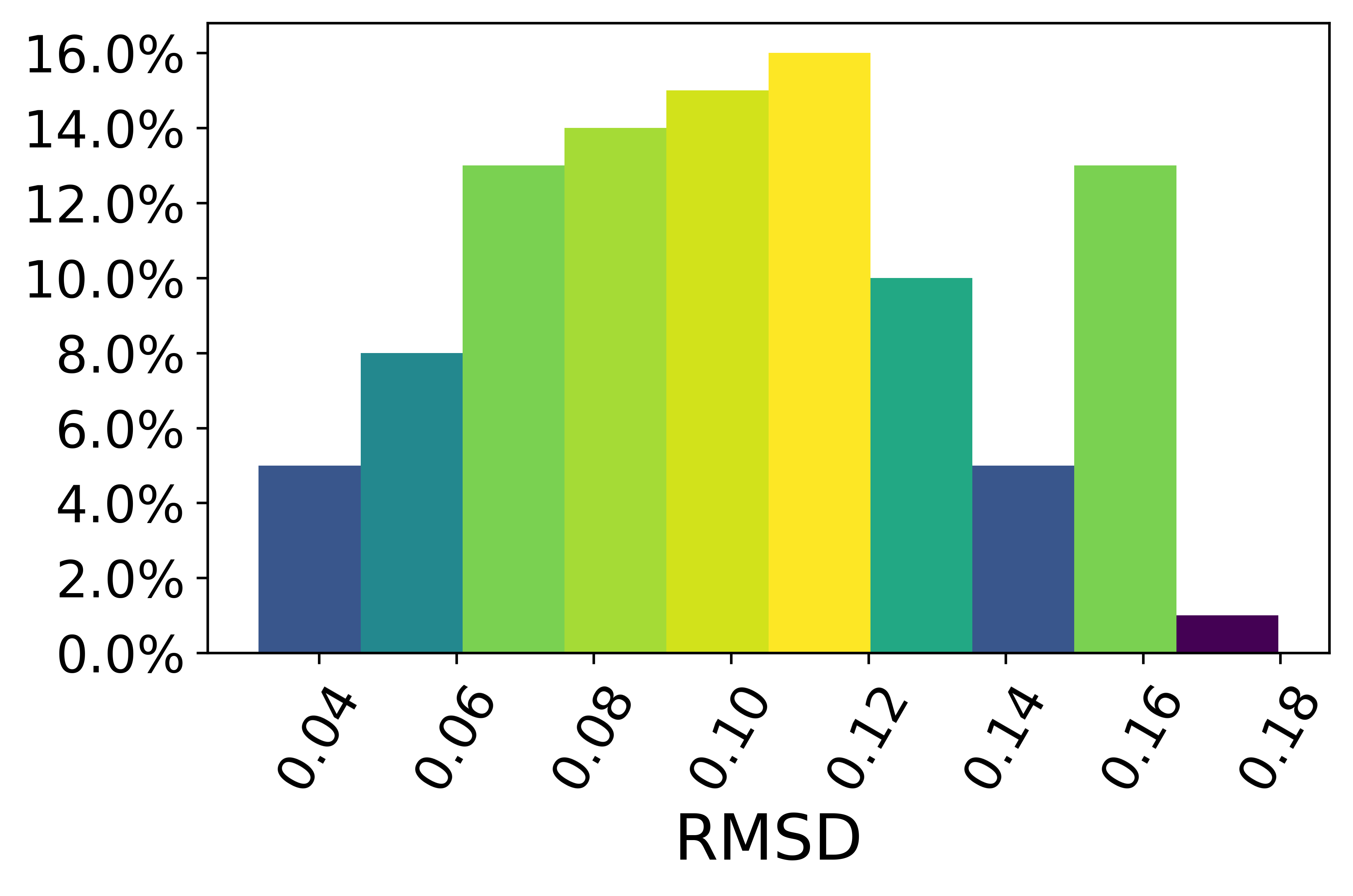}

	\caption{Histogram of the root-mean-square deviations (RMSD) from target fractional composition. The  RMSD were computed with \cref{eq:fitness_comp} derived from \cref{eq:fitness_N}
 }
	\label{fig:composition_rmsd}
\end{figure}

However, it is worth noting that the minimum never reaches 0 for all three figures. For an equiatomic 4-components system, the total fitness of the maximum-entropy configuration is equal to 0. The NESs training process does not always converge to this maximum-entropy configuration, thus introducing imperfections that can be characterized by evaluating the deviation from the target composition.  One hundred equiatomic \ce{Cu_{\alpha}Ni_{\beta}Co_{\gamma}Cr_{\zeta} } structures  with 256 atoms were generated  using the same NES model and the root-mean-square deviation (RMSD) from target fractional composition was computed as 

\begin{align}
\label{eq:fitness_comp}
RMSD =  \sqrt{\dfrac{\sum_{a} \left[c_a' - c_a \right]^2 }{ s}},
\end{align}
where $c_a$ and  $c_a'$ are target the fractional composition of the a-type  atom,  and the fractional composition of the a-type in the NESs, and $s$ is the number of atom-types. The result is plot in \cref{fig:composition_rmsd}. The RMSD values varies from 0.03 to 0.18, and fractional composition  associated with the minimum value is  \ce{Cu_{0.234} Ni_{0.258} Co_{0.258} Cr_{0.250}}. The true value in this example is \ce{Cu_{0.25} Ni_{0.25} Co_{0.25} Cr_{0.25}}. To this end, multiple structures should be optimized in parallel and one should select the structure with the highest score (low fitness function).

\subsection{ FCC  Multicomponent  alloys}

\begin{figure}[hbt!]
		\centering
		\includegraphics[width=0.8\linewidth]{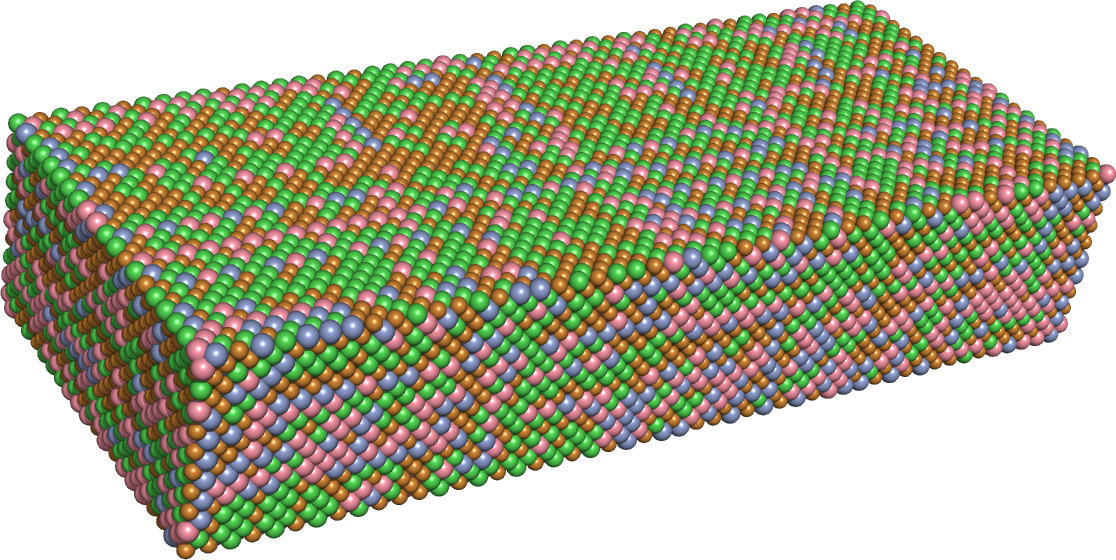}
	\caption{Example of \ce{Cu_{0.234} Ni_{0.320} Co_{0.226} Cr_{0.220}} NES structure (40,000 atoms). The model was trained on a 8-atoms input-structure and the generation was completed in 328 minutes.}.
	\label{fig:quaternay_structures}
\end{figure}

\begin{figure}[hbt!]
	\centering
	\includegraphics[width=0.75\linewidth]{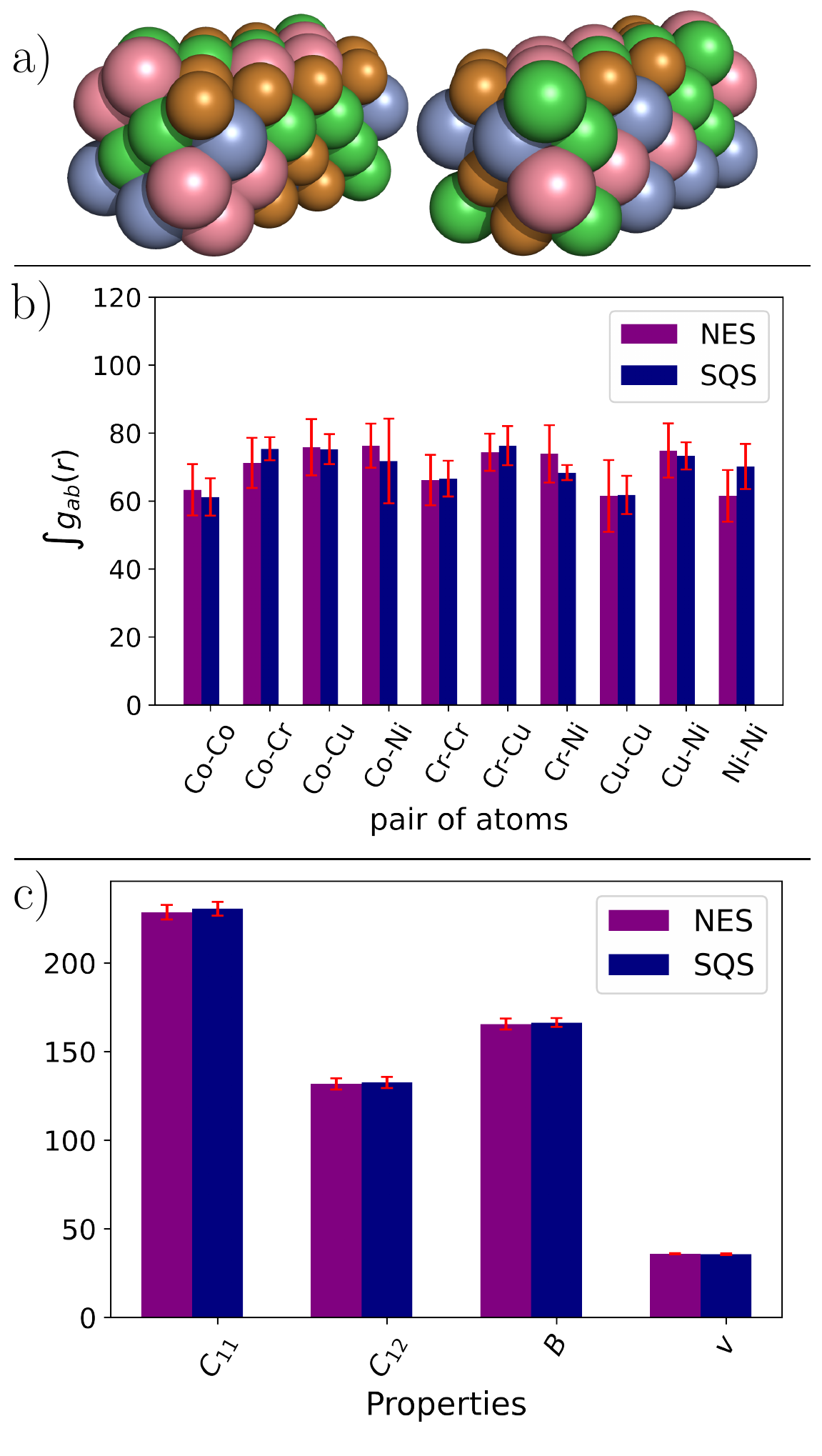}
	
	\caption{NES vs SQS. a) Selected 64-atoms equiatomic \ce{Cu_{\alpha}Ni_{\beta}Co_{\gamma}Cr_{\zeta} } NES structures.
	b) Comparison of partial PDF [\cref{eq:N_ab}]. Each  bar corresponds to the average over ten structures  of the area under the first peak of $g_{ab}(r)$.
	b) Selected computed properties. $C_{11}$ and $C_{12}$ are the Elastic constant (GPa), $B$ is the Bulk modulus (GPa), and $v$ is the Poisson ratio ($\times 100$). Properties were computed using the classical molecular dynamics}
	\label{fig:comparison}
\end{figure}

NESs generation was applied to build  equiatomic \ce{Cu_{\alpha}Ni_{\beta}Co_{\gamma}Cr_{\zeta} }  FCC  alloy structures, and the performance was then evaluated by generating a structure with  40,000  atoms (\cref{fig:quaternay_structures}). The model was trained with a   $2\times 2 \times 2$ cell (8 atoms) and the generation was completed in 328 minutes. In addition,  a number of 64-atoms structures  were generated and selected properties were compared to ICET-SQS \cite{ICET} (\cref{fig:comparison}). \Cref{fig:comparison}-a shows selected structures used for the comparison.
 
The partial PDFs of the structures are compared in \cref{fig:comparison}-b. Each bar corresponds to the average of the area under the first peak of the $g_{ab}(r)$ (coordination numbers). The standard deviations are plotted in red. The purple bars represent NES and the blue bars represent SQS. The analysis of the chart shows that NES is almost equivalent to SQS. Indeed, for each pair, the values of $g_{ab}(r)$ for NES are almost always within one standard deviation of the SQS values.

Second, the Elastic constants, Bulk modulus, and the Poisson's ratio were also investigated using classical Molecular Dynamics. These simulations were carried out using the LAMMPS molecular dynamics simulator \cite{plimpton1995fast} and an Embedded Atom Method (EAM) potential was used to define the inter-atomic interactions\cite{farkas2018model}. ICET-SQS and NES of equiatomic \ce{Cu_{\alpha}Ni_{\beta}Co_{\gamma}Cr_{\zeta} } was systematically deformed and the change in virial stress tensor was used to calculate the elastic constants. Each deformed structure was energetically minimized using the conjugate gradient algorithm\cite{polak1969note} before performing the stress calculations. All simulations were performed at 0 K. The bar chart of these properties is plotted in \cref{fig:comparison}-b. All of the calculated values from the NES structure is within one standard deviation of the SQS method.

\section{Conclusions}
\label{conclusion}
We introduce and utilize a neural evolution structures (NESs) generation methodology combining artificial neural networks (ANNs) and evolutionary algorithms (EAs) to generate High Entropy Alloys (HEAs). Our inverse design approach based on pair distribution functions and atomic properties dramatically reduces computational cost, allowing for the generation of very large structures with over 40,000 atoms in few hours.  The computing time is speed-up factor of  about  1000 with respect to the SQSs. Unlike the SQSs, the same model can be used to generate multiple structures with same fractional composition. A number of  NE structures have been using to compute selected properties such as the elastic constants, the bulk modulus, and the Poisson ratio, and the results are similar to those of structures generated with SQS.

\begin{acknowledgments}
KR, CVS, IT acknowledge support from NSERC. Work at NRC was carried out under the auspices of the AI4D and MCF Programs.

\end{acknowledgments}

\bibliography{biblio}

\end{document}